\documentclass[a4paper,12pt]{article}
\usepackage{amsfonts}
\usepackage{amsmath}
\usepackage{mdsymbol}

\usepackage{graphicx}

\usepackage{layaureo}
\newenvironment{proof}[1][Proof]{\noindent\textbf{#1.} }{\ \rule{0.5em}{0.5em}}

\begin{document}
\title{On Trade in Bilateral Oligopolies\\ with Altruistic and Spiteful Agents\thanks{%
We would like to thank Sayantan Ghosal for his helpful comments and
suggestions. The usual disclaimer applies.}}
\author{M Lombardi\thanks{Adam Smith Business School, University of Glasgow, Glasgow, G12 8QQ, UK.\newline E-mail:michele.lombardi@glasgow.ac.uk} \ and S Tonin\thanks{Durham University Business School, Durham University, Durham, DH1 3LB, UK.\newline E-mail:
simone.tonin@durham.ac.uk}}
\maketitle

\begin{abstract}
\noindent This paper studies the effects of altruism and spitefulness in a two-sided market in which agents behave strategically and trade according to the Shapley-Shubik mechanism. By assuming that altruistic agents have concerns for others on the opposite side of the market, it shows that agents always find advantageous to trade. However, they prefer to stay out of the market and consume their endowments when there are altruistic agents who have concerns for the welfare of those on the same side of the market, or when there are spiteful agents. These non-trade situations occur \textit{either} because the necessary first-order conditions for optimality are violated \textit{or} because agents' payoff functions are not concave.\\

\noindent \textit{JEL classification:} D43, D51.

\noindent \textit{Keywords:} Bilateral oligopoly, noncooperative oligopoly, Nash equilibrium, altruism and spitefulness.\newpage
\end{abstract}

\section{Introduction}
We often incorporate the preferences of others in our decision making. We do so because we intrinsically care about the welfare of other agents in the economy. In this paper, we continue a line of inquiry begun by Dubey and Shubik (1985) and Dufwenberg et al (2011) by investigating how altruism and spitefulness influence equilibrium outcomes in imperfectly competitive markets.\par
We confine ourselves to a class of strategic market games introduced by Gabszewicz and Michel (1997), known as bilateral oligopolies.\footnote{This basic model of trade has been also studied by Bloch and Ghosal (1997), Bloch and Ferrer (2001), Dickson and Hartley (2008), Amir and Bloch (2009), among others.} In this two-sided market model, agents act strategically and trade according to Shapley and Shubik (1977)'s mechanism (henceforth, Shapley-Shubik mechanism): They submit bids and offers to the mechanism and the price is determined by the ratio of total bids to total offers.\par
The cornerstone of bilateral oligopolies is the assumption that individual agents' behaviors are solely motivated by their personal concern. However, there is a considerable amount of both experimental and empirical evidence that individuals do not have independent preferences, in the sense that considerations of others influence individual behavior. This paper departs from the traditional assumption of independent preferences by assuming that agents act by considering both personal concern and concerns for the welfare of others.\par
By following the growing literature on behavioral economics that constructs theoretical models with altruistic/spiteful agents (e.g., Levine, 1998; Bourl\`{e}s et al, 2017), this paper also assumes that an agent, who has a concern for others, has an overall utility function which encompasses both his internal utility (that is, a classical utility function defined over his consumption set) and internal utilities of others weighted by preference parameters.\footnote{For excellent surveys see, for instance, Fehr and Gachter (2000) and Sobel (2005).}  Each parameter reflects the degree of importance that an agent puts on the welfare of another: positive under altruism, negative under spitefulness and zero under the classic assumption of independent preferences. We refer to this utility function as Edgeworth utility function (Edgeworth, 1881).\par
Dufwenberg et al (2011) show that concern for others does not affect equilibrium outcomes when markets are perfectly competitive and agents' preferences are represented by Edgeworth's utility functions. In particular, they find that agents behave \textit{as if} they had classical independent preferences at competitive equilibria. Dubey and Shubik (1985) reach exactly the same conclusion with a continuum of agents in a strategic market game. By contrast, we find that altruism and spitefulness affect the volume of trade in bilateral oligopolies and, more interestingly, that this type of preferences may shrink the volume of trade down to zero. This holds even when trade produces high internal utility gains - a requirement introduced by Bloch and Ferrer (2001) in response to the non-trade situations studied by Cordella and Gabszewicz (1998). These findings are not in line with the conclusions reached in auction settings (e.g., Levine, 1998; Sobel, 2010), according to which concern for others does not affect equilibrium outcomes.\par
When agents have altruistic concerns for others on the same side of the market, we provide an example of bilateral oligopoly satisfying the classical assumptions on utility functions, as well as an assumption of high internal gains from trade, for which the non-trade equilibrium is the unique (Nash) equilibrium.\footnote{A non-trade equilibrium, also called trivial equilibrium, always exists. To study trade equilibria, Bloch and Ferrer (2001) introduce an assumption on marginal utilities, similar to the Inada conditions, which implies high internal gains from trade.} The economic intuition for this negative result is as follows. A property of the Shapley-Shubik mechanism is that an offer by an agent produces internal utility losses for others on the same side of the market. This translates, \textit{ceteris paribus}, into a utility loss for a supplier who intends to maximize the welfare of others on the same side of the market. When such a loss is not compensated by a gain in his internal utility, the altruistic supplier would prefer to offer nothing to the market.\par
However, we also show that a trade equilibrium exists when agents have altruistic concerns for others on the opposite side of the market. We obtain this existence result under assumptions that are common in the bilateral oligopoly literature with independent preferences (e.g., Bloch and Ferrer, 2001). This positive result can be explained as follows. When agents have independent preferences and there are high gains from trade, there exists a trade equilibrium. A property of the Shapley-Shubik mechanism is that an offer by an agent produces internal utility gains for others on the opposite side of the market. This translates, \textit{ceteris paribus}, into utility gains for agents who have altruistic concerns for others on the opposite side of the market. Therefore, this type of altruism strengthens the incentives to trade and does not upset the conclusions drawn under the assumption of independent preferences. Our new proof of existence can be used to generalize Bloch and Ferrer (2001)'s existence result.\par
While there are many settings where agents are altruistic, there are also situations in which agents aim to outdo other agents in order to improve their own standing. For this reason, we also study bilateral oligopolies with spiteful agents, who are interested in minimizing the welfare of others as well as in maximizing their internal utilities. In this setting, we provide examples of bilateral oligopolies, satisfying the classical assumptions on utility functions, as well as the assumption of high internal gains from trade, for which the non-trade equilibrium is the unique equilibrium. In particular, we explore the effects of spiteful concerns both for agents on the opposite side of the market and for those on the same side of the market. Although spitefulness is detrimental to the existence of trade equilibria, we find that these negative effects are caused by two distinct factors.\par
In a setting where agents aim to minimize the welfare of others on the opposite side of the market, the non-existence of trade equilibria stems from the fact that in the Shapley-Shubik mechanism an offer by an agent produces internal utility gains for others on the opposite side of the market. This translates, \textit{ceteris paribus}, into a utility loss for a supplier who intend to minimize the welfare of others on the opposite of the market. When such a loss is not compensated by a gain in his internal utility, a spiteful supplier would prefer to offer nothing to the market. By contrast, in a setting where agents aim to minimize the welfare of others on the same side of the market, we report that the non-existence is due to the non-concavity of the payoff functions.\par
Section 2 describes the theoretical framework and outlines the bilateral oligopoly model, with results presented in Section 3. Section 4 concludes. The appendix contains the proof of existence.

\section{Mathematical model}
We consider exchange economies with two types of agents, labelled 1 and 2,
and two (perfectly divisible) commodities, labelled $x$ and $y$. The set of
agents is $I=I^{1}\cup I^{2}$, where $I^{t}$ is the finite set of agents of type $t=1,2$. An agent $i$ is of type 1 when he is endowed with $x_{i}^{0}>0$ units of commodity $x$ but no unit of commodity $y$. Similarly, an agent $i$ is of type 2 when he is endowed with $y_{i}^{0}>0$ units of commodity $y$ but no unit of commodity $x$. Therefore, agents of different types are on different sides of the market as they hold different commodities. We make the following assumption throughout the paper.

\bigskip

\noindent \textbf{Assumption 1} There are at least two agents for each type.

\bigskip

Agent $i$'s bundle $\left( x_{i},y_{i}\right) $ is a (non-negative)
two-dimensional vector describing how much of each commodity he consumes. An
allocation $\left( x,y\right) =\left( x_{i},y_{i}\right) _{i\in I}$ is a
list of bundles. Each agent $i$ maximizes the utility function: 
\begin{equation*}
V_{i}\left( x,y\right) =u_{i}(x_{i},y_{i})+\sum_{j\neq i}\mathcal{\gamma }%
_{i}^{j}u_{j}(x_{j},y_{j})\text{,}
\end{equation*}%
where $-1\leq \gamma _{i}^{j}\leq 1$ for each agent $j\neq i$.\footnote{%
Edgeworth (1881) calls the parameter $\gamma _{i}^{j}\geq 0$ coefficient of
effective sympathy. See Levine (1988) for a discussion on the different
interpretations of $\gamma _{i}^{j}$.} Agent $i$'s utility function
embodies a private and a social component. The private component is
represented by the internal utility $u_{i}$ that depends on commodities that
go directly to him. The social component is instead represented by the
weighted sum of the internal utilities of other agents. Each parameter $%
\mathcal{\gamma }_{i}^{j}$ is the weight that agent $i$ put on agent $j$'s
utility, positive under altruism, negative under spitefulness, and zero
under independent preferences. Let $\gamma
_{i}=( \gamma _{i}^{j}) _{j\in I} $, with $\gamma _{i}^{i}=1$, denote agent $%
i$'s preference parameters. The following classical assumption on
internal utilities is made.

\bigskip

\noindent \textbf{Assumption 2} For each agent $i\in I$, the internal
utility $u_{i}$ is continuous, continuously differentiable,\footnote{%
Differentiability should be implicitly understood to include the case of
infinite partial derivatives along the boundary of the consumption set (see
Kreps (2012), p. 58.)} strictly increasing and concave.

\bigskip

By following the approach developed by Bloch and Ferrer (2001), we consider bilateral oligopolies in which there are high internal gains from trade, i.e., high gains from trade with respect to internal utility functions.\footnote{Bloch and Ferrer (2001) assume that the restriction on marginal utilities holds for both commodities and for all agents.} This is captured by the following assumption.

\bigskip \noindent \textbf{Assumption 3} There exists an agent $i\in I^{1}$
such that $\frac{\partial u_{i}(x_{i},0)}{\partial y_{i}}=\infty $, for each 
$x_{i}>0$, or an agent $i\in I^{2}$ such that $\frac{\partial u_{i}(0,y_{i})%
}{\partial x_{i}}=\infty $, for each $y_{i}>0$. 

\bigskip

Since the set of agents will remain fixed, an exchange economy with altruistic/spiteful agents is denoted by $(u,\gamma ,w)$ where $u=\left( u_{i}\right) _{i\in I}$
is the profile of internal utilities, $\gamma=\left( \gamma _{i}\right) _{i\in
I}$ is the profile of agents' profile of preference parameters, and $w=(x_{i}^{0},y_{i}^{0})_{i\in I}$ is the endowment profile. An exchange economy in which agents have independent preferences is simply denoted by $(u,w)$.

In our model, agents behave strategically: each agent offers a quantity of
his endowment to the market. The strategy spaces are thus given by:%
\begin{equation*}
S_{i}=\left\{ a_{i}:0\leq a_{i}\leq x_{i}^{0}\right\} \text{,}\hspace*{4mm}%
\hspace*{4mm}\text{for each }i\in I^{1}\text{,}
\end{equation*}%
\begin{equation*}
S_{i}=\left\{ b_{i}:0\leq b_{i}\leq y_{i}^{0}\right\} \text{,}\hspace*{4mm}%
\hspace*{4mm}\text{for each }i\in I^{2}\text{.}
\end{equation*}%
We write $(a,b)$ for the profile of offers $((a_{i})_{i\in
I^{1}},(b_{i})_{i\in I^{2}})$ and $S$ for $\tprod\limits_{i\in I}S_{i}$.
Clearly, $(a,b)$ is an element of $S$. Additionally, $(a_{-i},b)$ is an
element of $\tprod\limits_{j\neq i}S_{j}$, with $i\in I^{1}$, and $%
(a,b_{-i}) $ is an element of $\tprod\limits_{j\neq i}S_{j}$, with $i\in
I^{2}$. For any profile of offers $(a,b)\in S$, the bundles assigned to
agents are given by the following allocation rule:%
\begin{equation}
\left( x_{i}\left( a,b\right) ,y_{i}\left( a,b\right) \right) =\left(
x_{i}^{0}-a_{i},a_{i}\frac{B}{A}\right) \text{,}\hspace*{4mm}\hspace*{4mm}%
\text{for each }i\in I^{1}\text{,}  \label{allocation1}
\end{equation}%
\begin{equation}
\left( x_{i}\left( a,b\right) ,y_{i}\left( a,b\right) \right) =\left( b_{i}%
\frac{A}{B},y_{i}^{0}-b_{i}\right) \text{,}\hspace*{4mm}\hspace*{4mm}\text{%
for each }i\in I^{2}\text{,}  \label{allocation2}
\end{equation}%
if $A=\dsum_{i\in I^1}a_{i}>0$ and $B=\dsum_{i\in I^2}b_{i}>0$. Otherwise,
each agent consumes only his own endowment. The ratio $B/A$ is the price of
commodity $x$, whereas the price of $y$ is normalized to 1. The allocation
generated by $\left( a,b\right) $ is denoted by $\left( x\left( a,b\right)
,y\left( a,b\right) \right) $.

The above allocation rule combined with $(S,u,\gamma )$ defines a bilateral
oligopoly with altruistic and spiteful agents, which is denoted by $\Gamma (\gamma )$. We write $\Gamma $ to denote a bilateral oligopoly where agents have
independent preferences. We adopt the solution concept of (Nash) equilibrium.

\bigskip

\noindent \textbf{Definition 1 }An equilibrium for $\Gamma \left(
\gamma \right) $ is a profile of offers $(\hat{a},\hat{b})$ such that:

\begin{itemize}
\item For each agent $i\in I^{1}$, $\hat{a}_{i}\in S_{i}$ maximizes: 
\begin{equation*}
V_{i}\left( x\left( \left( a_{i},\hat{a}_{-i}\right) ,\hat{b}\right)
,y\left( \left( a_{i},\hat{a}_{-i}\right) ,\hat{b}\right) \right) \text{.}
\end{equation*}

\item For each agent $i\in I^{2}$, $\hat{b}_{i}\in S_{i}$ maximizes:%
\begin{equation*}
V_{i}\left( x\left( \hat{a},\left( b_{i},\hat{b}_{-i}\right) \right)
,y\left( \hat{a},\left( b_{i},\hat{b}_{-i}\right) \right) \right) \text{.}
\end{equation*}
\end{itemize}

\bigskip

Let $\left( \hat{a},\hat{b}\right) $ be an equilibrium for $\Gamma \left( \gamma
\right) $. The profile $\left( \hat{a},\hat{b}\right) $ is a non-trade
equilibrium for $\Gamma \left( \gamma \right) $ if $\hat{A}=0$ and $\hat{B}%
=0 $.  The non-trade equilibrium always exists. The profile $%
\left(\hat{a},\hat{b}\right) $ is a trade equilibrium for $\Gamma \left(
\gamma \right) $ if $\hat{A}>0$ and $\hat{B}>0$.

\section{Altruistic agents}
We start our analysis by considering altruistic agents who are characterized by positive preference parameters. They aim to maximize their internal utilities as well as the welfare of other agents.\par
Let us start considering the case where agents have altruistic concerns for others on the opposite side of the market. In such a context, a trade equilibrium always exists as shown by the following theorem.

\bigskip

\noindent \textbf{Theorem} Let $(u,\gamma,w)$ be an exchange economy satisfying Assumptions 1-3. For each agent $i\in I^{t}$, let $\gamma_i$ be such that $\gamma _{i}^{j}=0$ for each $j\in I^{t}\backslash \left\{ i\right\}$ and $\gamma _{i}^{j}\geq 0 $ for each $j\in I\backslash I^{t}$, for each $t=1,2$. Then, there exists a trade 
equilibrium for $\Gamma \left( \gamma \right) $.

\bigskip

The proof can be found in the appendix and it adopts techniques which are similar to the ones used by Dubey and Shubik (1978) and Bloch and Ferrer (2001). The main novelty is the way in which the Kuhn-Tucker Theorem is used to show that there is trade in equilibrium. Furthermore, we can obtain the following corollary which
generalizes the existence result obtained by Bloch and Ferrer (2001) for
bilateral oligopolies with independent preferences. \bigskip

\noindent \textbf{Corollary} Let $(u,w)$ be an exchange economy satisfying Assumptions 1-3. Then, there exists a trade equilibrium for $\Gamma$.

\bigskip

The proof follows immediately from the theorem when $\gamma_i $ is set equal to zero for all agents.

We now turn to the case where agents have altruistic concerns for others on the same side of the market. In sharp contrast to the previous result, we show, by means of an example, that a trade equilibrium may fail to exist, though agents' internal utilities satisfy Assumptions 2-3. This non-existence result is due to the negativity of the necessary first-order conditions for optimality.
 \bigskip

\noindent \textbf{Example 1}. Consider an exchange economy with
four agents having the following utility functions and endowments: 
\begin{align*}
& V_{1}(x,y)=\frac{2}{3}x_{1}+y_{1}+\frac{1}{2}\left( \frac{2}{3}%
x_{2}+2y_{2}\right) \text{ and }(x_{1}^{0},y_{1}^{0})=(4,0), \\
& V_{2}(x,y)=\frac{2}{3}x_{2}+2y_{2}\text{ and }(x_{2}^{0},y_{2}^{0})=(4,0)%
\text{,} \\
& V_{i}(x,y)=\sqrt{x_{i}}+y_{i}\text{ and }(x_{i}^{0},y_{i}^{0})=(0,4),\text{
for }i=3,4.
\end{align*}%
Note that agents' internal utilities satisfy Assumptions 2-3. Also, note that agent 1 has an altruistic concern for agent 2. By checking the necessary first-order conditions for optimality, it is straightforward to verify that the non-trade equilibrium is the unique equilibrium of the $\Gamma (\gamma )$ associated with the aforesaid exchange economy.

\bigskip

The reason behind this non-existence result can be explained as follows.
From the allocation rule (\ref{allocation1}), the final quantity of the
commodity $y$ assigned to agent 2 depends negatively on the quantity of $x$
offered by agent 1. Therefore, if agent 1's gain from consuming additional
units of commodity $y$ does not outweigh his loss from a decrease in agent
2's consumption of $y$, then agent 1 maximizes his payoff by reducing his
offer $a_{1}$ to zero. This can be seen by considering the
derivative of the payoff function of agent 1 with respect to his offer $%
a_{1} $, which can be stated as follows:%
$$
\frac{\partial V_{1}}{\partial a_{1}}=-\frac{\partial u_{1}}{\partial x_{1}}+%
\frac{\partial u_{1}}{\partial y_{1}}\frac{B}{A^{2}}a_{2}-\gamma
_{1}^{2}\left( \frac{\partial u_{2}}{\partial y_{2}}\frac{B}{A^{2}}%
a_{2}\right) \text{.}  $$
By substituting the marginal utilities of agents 1 and 2 as well as
agent 1's preference parameter $\gamma _{1}^{2}=\frac{1}{2}$ of Example 1,
one can easily verify that the above derivative is negative and then agent
1's best strategy is $a_{1}=0$. It is well know that in the bilateral
oligopoly model we have a trade equilibrium only when there are two agents
offering each commodity.\footnote{%
When $a_{1}=0$, the allocation rule (\ref{allocation1}) implies that agent 2
can increase his utility by decreasing his offers of commodity $x$ because
for any $a_{2}$ he gets all the amount $B$ of commodity $y$ offered in the
market. In such a case the agent 2's payoff function is not continuous on $%
S_{2}$ and he does not have a best strategy.} Since in our example agent 1's
best offer is always nil, the non-trade equilibrium is the unique equilibrium.

\section{Spiteful agents}
We now consider bilateral oligopolies with spiteful agents. Such agents are characterized by negative preference parameters and they aim to minimize the welfare of other agents in the economy. Although there are high internal gains from trade, we show, by means of examples, that a trade equilibrium may fail to exist when there are spiteful agents.\par
We first turn to the case where agents have spiteful concerns for others on the opposite side of the marker. In such a case, the non-existence of a trade equilibrium is due to the negativity of the necessary first-order conditions for optimality.\par 
\bigskip

\noindent \textbf{Example 2}. Consider an exchange economy with
four agents having the following utility functions and endowments: 
\begin{align*}
& V_{i}(x,y)=\frac{2}{3}x_{i}+y_{i}-\frac{1}{2}\left( \sqrt{x_{3}}%
+y_{3}\right) -\frac{1}{2}\left( \sqrt{x_{4}}+y_{4}\right) \text{ and }%
(x_{i}^{0},y_{i}^{0})=(4,0),\text{ for }i=1,2, \\
& V_{i}(x,y)=\sqrt{x_{i}}+y_{i}\text{ and }(x_{i}^{0},y_{i}^{0})=(0,4),\text{
for }i=3,4.
\end{align*}%
Note that agents' internal utilities satisfy Assumptions 2-3. Also, note that agents $1$ and $2$ have spiteful concerns for agents $3$ and $4$. It is possible to verify that the non-trade equilibrium is the unique equilibrium of the $\Gamma (\gamma )$ associated with the aforesaid exchange economy.%
\footnote{This can be verified by solving the necessary first order condition for
optimality with any computer algebra system.}

\bigskip

The intuition behind this non-existence result can be explained as follows.
From the allocation rule (\ref{allocation2}) the final quantities of
commodity $x$ assigned to agents 3 and 4 depend positively on the quantity $%
x $ offered by agents 1 and 2. Therefore, if agent $1$'s gain from
consuming additional units of commodity $y$ does not outweigh his loss
from an increase in agents $3$ and $4$' consumption of commodity $x$, then
agent 1 may maximizes his payoff by reducing his offer $a_{1}$ to zero.
This can also be seen by considering the derivative of the payoff function
of agent $1 $ with respect to his offer, which can stated as follows: 
\begin{equation*}
\frac{\partial V_{1}}{\partial a_{1}}=-\frac{\partial u_{1}}{\partial x_{1}}+%
\frac{\partial u_{1}}{\partial y_{1}}\frac{B}{A^{2}}a_{2}+\gamma
_{1}^{3}\left( \frac{\partial u_{3}}{\partial x_{3}}\frac{b_{3}}{B}\right)
+\gamma _{1}^{4}\left( \frac{\partial u_{4}}{\partial x_{4}}\frac{b_{4}}{B}%
\right) \text{.}
\end{equation*}%
Heuristically, if $\gamma _{1}^{3}$ and $\gamma _{1}^{4}$
are high enough, then agent 1's best strategy is $a_{1}=0$.\footnote{%
Recall that $\gamma _{i}^{j}<0$ under spitefulness.} The same argument applies to agent 2. We thus conclude that there is no strategy profile that can satisfy the
necessary first-order conditions for optimality, and so the non-trade equilibrium is the unique equilibrium of $\Gamma \left( \gamma \right)$.\par

\bigskip
We finally turn to the case where agents have spiteful concerns for others on the same side of the market. In such a case, the non-existence of a trade equilibrium is due to the non-concavity of payoff functions. 
\bigskip

\noindent \textbf{Example 3}. Consider an exchange economy with
four agents having the following utility functions and endowments: 
\begin{align*}
&V_{1}(x,y)=\frac{2}{3}x_{1}+y_{1}-\frac{1}{2}\left(x_{2}-y_{2}^{-2}\right) 
\text{ and } (x_{1}^{0},y_1^{0})=(4,0), \\
&V_{2}(x,y)=x_{2}-y_{2}^{-2}\text{ and }(x_{2}^{0},y_2^{0})=(4,0)\text{,} \\
&V_{i}(x,y)=\sqrt{x_{i}}+y_{i}\text{ and }(x_{i}^{0},y_{i}^{0})=(0,4),\text{
for } i=3,4.
\end{align*}
Note that the internal utilities of agents satisfy Assumptions 2-3.%
\footnote{The internal utility function of agent 2 is not defined on the boundary of
the consumption set. This does not affect the current analysis but should be
kept in mind. Examples with an internal utility function
defined also on the boundary can be found, e.g., $V_{2}(x,y)=x_{2}-(\frac{1}{%
10}+y_{2})^{-2}$.} Also note that only agent $1$ has spiteful concern for agent $2$. 
It is possible to verify that the non-trade equilibrium is the unique equilibrium of the $\Gamma (\gamma )$ associated with the aforesaid exchange economy.

\bigskip

In the example the strategy profile $(\hat{a}_{1},%
\hat{a}_{2},\hat{b}_{3},\hat{b}_{4})=(3.097,3.673,0.460,0.460)$ is the
unique solution to the necessary first-order conditions for optimality of
all agents.\footnote{This can be verified with any computer algebra system} However, this profile does not correspond to an equilibrium because agent 1 can find a unilateral profitable deviation. Indeed, the figure below shows the shape of agent 1's payoff
function when other agents offer $(\hat{a}_{2},\hat{b}_{3},\hat{b}_{4})$. It
is immediate to see that agent 1's payoff function is convex and $\hat{a}%
_{1} $ corresponds to a minimum point. Therefore, $(\hat{a}_{1},\hat{a}_{2},%
\hat{b}_{3},\hat{b}_{4})$ cannot be an equilibrium, and so the non-trade equilibrium is the unique equilibrium of $\Gamma(\gamma)$.

\begin{center}\includegraphics[scale=0.45]{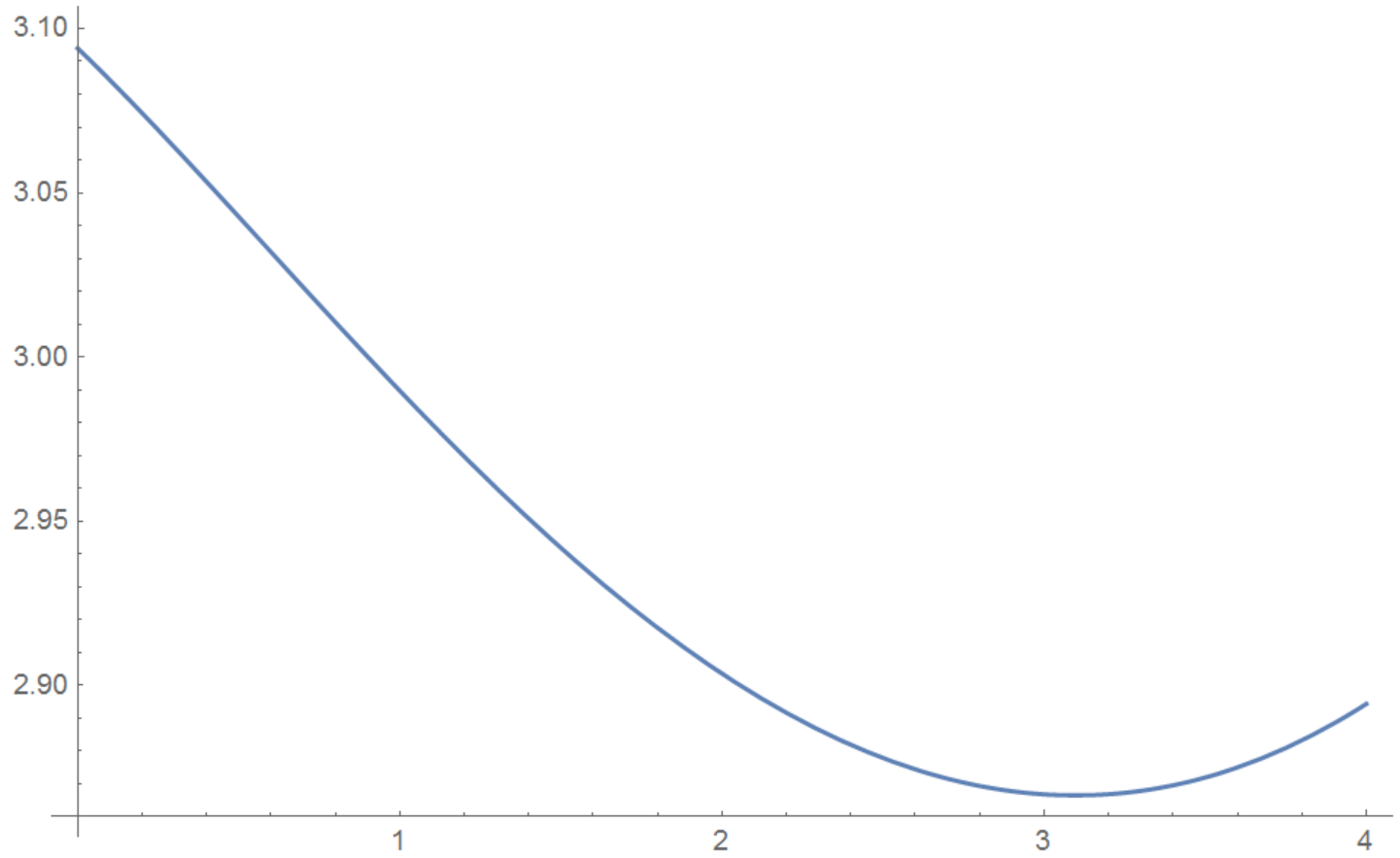}\end{center}

\section{Concluding remarks}
In this paper, we study the effects of altruism and spitefulness in a bilateral oligopoly. We prove that a trade equilibrium exists when agents have altruistic concerns for others on the opposite side of the market. The intuition behind this positive result is that incentives to trade are strengthened under this configuration of altruistic concerns. By contrast, we show, by means of examples, that the non-trade equilibrium is the unique equilibrium in all other cases analyzed. These negative results are caused by the negativity of the necessary first-order conditions for optimality (as in Examples 1 and 2) and by the non-concavity of payoff functions (as in Example 3).\par
Before closing the paper, we wish to call attention to two points. First, we confine ourself to bilateral oligopolies with corner endowments in one commodity. We do not know whether our negative results extend to models with interior endowments. This is left for further research. Second, as in Dubey and Shubik (1985), one can show that the equilibrium of a bilateral oligopoly corresponds to the competitive equilibrium when there is a continuum of altruistic and spiteful agents. However, we still do not know whether the equilibrium of a bilateral oligopoly converges to the competitive equilibrium when the underlying exchange economy -with altruistic and spiteful agents- is replicated. The reason for this is that the standard convergence results do not apply (e.g., Lemma 4 of Dubey and Shubik, 1978), even when there exists a trade equilibrium. This is a fruitful research area for future works.

\appendix
\section{Appendix}
The proof of the existence theorem is based on three lemmas which require
the following preliminary result.

\bigskip

\noindent \textbf{Proposition} Let Assumption 2 hold. For each agent $i\in I^{t}$, let $\gamma_i$ be such that $\gamma _{i}^{j}=0$ for each $j\in
I^{t}\backslash \left\{ i\right\}$ and $\gamma _{i}^{j}\geq 0$, for each $%
j\in I\backslash I^{t}$, for each $t=1,2$. Then, the utility function $V_i$ is continuous,
monotone, and concave, for each $i\in I$.

\bigskip

\begin{proof}
Consider the utility function $V_{i}$ of an agent $i$. It is straightforward
to verify that $V_{i}$ is continuous, monotone, and concave as it is a sum
of continuous, monotone, and concave internal utility functions by
Assumption 2.
\end{proof}

\bigskip 

Following Dubey and Shubik (1978), in order to prove the existence of a (Nash) equilibrium, we introduce a perturbed game $\Gamma^\epsilon(\gamma)$, with $%
\epsilon\in(0,1]$. This is a game defined as $\Gamma(\gamma)$ with the only
exception that in the allocation rules (\ref{allocation1}) and (\ref%
{allocation2}) the ratio $\frac{B}{A}$ is replaced by $\frac{B+\epsilon}{%
A+\epsilon}$, i.e., the price of commodity $x$ becomes $\frac{B+\epsilon}{%
A+\epsilon}$. The interpretation is that an outside agency places a fixed
bid of $\epsilon$ and a fixed offer of $\epsilon$ in the trading post. This
does not change the strategy sets of agents, but does affect the prices, the
final holdings, and the payoffs. We denote by $(\hat{a}^\epsilon,\hat{b}%
^\epsilon)$ an equilibrium of the perturbed game $\Gamma^\epsilon(\gamma)$.

In the first lemma, we prove the existence of an equilibrium in the perturbed game.

\bigskip

\noindent \textbf{Lemma 1} Let Assumptions 1-3 hold. For each $\epsilon\in(0,1]$, there exists an equilibrium for 
$\Gamma^\epsilon(\gamma)$.

\bigskip

\begin{proof}
Consider an agent $i$ of type 1 and fix the strategies $(a_{-i},b)$ for all
other agents. In the perturbed game the payoff function $%
V_{i}(x(a,b),y(a,b)) $ is continuous as $\frac{B+\epsilon }{A+\epsilon }$ is
positive for each $\epsilon \in (0,1]$. Let 
\begin{equation*}
\phi _{i}(a_{-i},b)\in \underset{a_i\in S_i}{\arg \!\max }\ V_{i}(x(a_{-i},b),y(a_{-i},b))\end{equation*}
be the best response correspondence of the agent $i$. By the Weierstrass
Theorem, the best response correspondence $\phi _{i}$ is non-empty. We now
show that the correspondence $\phi _{i}$ has convex-valued. Suppose that
there are two feasible strategies $a_{i}^{\prime }$ and $a_{i}^{\prime
\prime }$ which belong to $\phi _{i}(a_{-i},b)$. We need to prove that $%
\tilde{a}_{i}=\delta a_{i}^{\prime }+(1-\delta )a_{i}^{\prime \prime }$,
with $\delta \in (0,1)$, belongs to $\phi _{i}(a_{-i},b)$. Since the
strategies $(a_{-i},b)$ are fixed, let us consider $(x(a,b),y(a,b))$ as
functions of $a_{i}$, i.e., $(x(a_{i}),y(a_{i}))$. Let $(x^{\prime
},y^{\prime })=(x(a_{i}^{\prime }),y(a_{i}^{\prime }))$, $(x^{\prime \prime
},y^{\prime \prime })=(x(a_{i}^{\prime \prime }),y(a_{i}^{\prime \prime }))$%
, and $(\tilde{x},\tilde{y})=\delta (x^{\prime },y^{\prime })+(1-\delta
)(x^{\prime \prime },y^{\prime \prime })$. Since the utility function $V_{i}$
is concave, by the Proposition, 
\begin{equation*}
V_{i}(\tilde{x},\tilde{y})\geq \delta V_{i}(x^{\prime },y^{\prime
})+(1-\delta )V_{i}(x^{\prime \prime },y^{\prime \prime })=V_{i}(x^{\prime
},y^{\prime }).
\end{equation*}%
From the allocation rules (\ref{allocation1}) and (\ref{allocation2}), we
have that $x_{i}(\tilde{a}_{i})=\tilde{x}_{i}$, as $x_{i}(a_{i})$ is linear; 
$y_{i}(\tilde{a}_{i})\geq \tilde{y}_{i}$, as $y_{i}(a_{i})$ is concave; and $%
x_{j}(\tilde{a}_{i})=\tilde{x}_{j}$, $y_{j}(\tilde{a}_{i})=\tilde{y}_{j}$,
as $x_{j}(a_{i})$ and $y_{j}(a_{i})$ are linear, for each $j\in I^{2}$. But
then, 
\begin{equation*}
V_{i}(x(\tilde{a}_{i}),y(\tilde{a}_{i}))\geq V_{i}(\tilde{x},\tilde{y}%
)=V_{i}(x^{\prime },y^{\prime })=V_{i}(x(a^{\prime }),y(a^{\prime })),
\end{equation*}%
as $V_{i}$ is monotone. Thus, $\tilde{a}_{i}$ maximizes agent $i$'s payoff
function and then it belongs to $\phi _{i}(a_{-i},b)$. Therefore, the
correspondence $\phi _{i}$ has convex-valued. Furthermore, by the Berge
Maximum Theorem, $\phi _{i}$ is an upper hemicontinuous correspondence. If
we consider an agent $i\in I^{2}$, then the previous argument leads, \textit{%
mutatis mutandis}, to the same result and $\phi _{i}(a,b_{-i})$ is a
non-empty, convex-valued, upper hemicontinuous correspondence. As we are looking for a
fixed point in the strategy space $S$, let's consider $\phi
_{i}:S\rightarrow S_{i}$. Let $\Phi :S\rightarrow S$ such that $\Phi
(S)=\prod_{i\in I}\phi _{i}(S)$. The correspondence $\Phi $ is a
convex-valued and upper hemicontinuous because it is a
product of convex-valued upper hemicontinuous correspondences. Moreover, $S$
is a compact and convex set. Therefore, by the Kakutani Fixed Point Theorem,
there exists a fixed point $(\hat{a}^{\epsilon },\hat{b}^{\epsilon })$ of $%
\Phi $, which is an equilibrium of the perturbed game $\Gamma ^{\epsilon }$.
\end{proof}

\bigskip

In the next lemma, we prove that the price of commodity $x$ is finite and
bounded away from zero at an equilibrium of any perturbed game.

\bigskip

\noindent \textbf{Lemma 2} At an equilibrium of the perturbed game $\Gamma^\epsilon(\gamma)$, $(\hat{a}^\epsilon,%
\hat{b}^\epsilon)$, there exist two positive constants $C$ and $D$,
independent from $\epsilon$, such that 
\begin{equation*}
C<\frac{\hat{B}^\epsilon+\epsilon}{\hat{A}^\epsilon+\epsilon}<D,
\end{equation*}
for each $\epsilon\in(0,1]$.

\bigskip

\begin{proof}
It is straightforward to see that the proof provided by Dubey and Shubik
(1978) still holds. To establish the existence of $C$, consider an agent $i$
of type 2. Following the same steps adopted by Dubey and Shubik (1978),
after having applied the Uniform Monotonicity Lemma, we still obtain the
following relationship on internal utility functions (see p. 10 in Dubey and
Shubik (1978)), 
\begin{equation*}
u_{i}(x_{i}(\Delta ),y_{i}(\Delta ))>u_{i}(x_{i}(\hat{a}^{\epsilon },\hat{b}%
^{\epsilon }),y_{i}(\hat{a}^{\epsilon },\hat{b}^{\epsilon })).
\end{equation*}%
The parameter $\Delta $ is a feasible increase in agent $i$ strategy and $%
(x_{i}(\Delta ),y_{i}(\Delta ))$ is the new corresponding bundle. Note that 
\begin{equation*}
\sum_{j\in I^{1}}u_{j}(x_{j}(\Delta ),y_{j}(\Delta ))>\sum_{j\in
I^{1}}u_{j}(x_{j}(\hat{a}^{\epsilon },\hat{b}^{\epsilon }),y_{j}(\hat{a}%
^{\epsilon },\hat{b}^{\epsilon }))
\end{equation*}
as $u_{j}$ is increasing in $y$, for each $j\in I^{1}$, by Assumption 2. From the two previous inequalities and since $\gamma _{i}^{j}=0$, for each $j\in
I^{2}\backslash \left\{ i\right\} $, and $\gamma _{i}^{j}\geq 0$, for each $%
j\in I^{1}$, we obtain that
\begin{equation*}
V_{i}(x(\Delta ),y(\Delta ))>V_{i}(x(\hat{a}^{\epsilon },\hat{b}^{\epsilon
}),y(\hat{a}^{\epsilon },\hat{b}^{\epsilon })).
\end{equation*}%
Since $(\hat{a}^{\epsilon },\hat{b}^{\epsilon })$ is an equilibrium, we obtain the
same contradiction of Dubey and Shubik (1978). By following their steps, we can then show that $\frac{\hat{B}^{\epsilon }+\epsilon }{\hat{A}%
^{\epsilon }+\epsilon }>C$. To establish the existence of $D$, consider an
agent $i$ of type 1. Then, the previous argument leads, \textit{mutatis
mutandis}, to $\frac{\hat{B}^{\epsilon }+\epsilon }{\hat{A}^{\epsilon
}+\epsilon }<D$.
\end{proof}

\bigskip

In the next lemma we use the Kuhn-Tucker Theorem to show that the agent
satisfying Assumption 3 places a positive offer at an equilibrium of any perturbed
game. This lemma is crucial to prove that there exists a trade equilibrium for $\Gamma(\gamma)$.

\bigskip

\noindent \textbf{Lemma 3} At an equilibrium of the perturbed game $\Gamma^\epsilon(\gamma)$, $(\hat{a}^\epsilon,%
\hat{b}^\epsilon)$, there exists two positive constants $\alpha$ and $\beta$%
, independent from $\epsilon$, such that 
\begin{equation*}
\alpha\leq \hat{A}^\epsilon\quad\mbox{or}\quad \beta\leq \hat{B}^\epsilon,
\end{equation*}
for each $\epsilon\in(0,1]$.

\bigskip

\begin{proof}
Let $(\hat{a}^\epsilon,\hat{b}^\epsilon)$ be an equilibrium of the perturbed game $%
\Gamma^\epsilon(\gamma)$. We first consider the case in which there exists
an agent $i\in I^1$ who satisfies Assumption 3. Then, $\hat{a}_i^\epsilon$
solves the following maximization problem 
\begin{equation*}  \label{max}
\begin{aligned} &\underset{a_i}{\text{max}}& & V_{i}(x((
a_i,\hat{a}_{-i}^\epsilon),\hat{b}^\epsilon)
,y((a_i,\hat{a}_{-i}^\epsilon),\hat{b}^\epsilon)),\\ & \text{subject to}& &
a_i\leq x_i^0,& &(i)\\ & & & -a_i \leq 0.& &(ii) \end{aligned}
\end{equation*}
By the Kuhn-Tucker Theorem, there exist non-negative multipliers $\hat{%
\lambda}_i$ and $\hat{\mu}_i$ such that 
\begin{align}  \label{focq1}
&\left.\frac{\partial V_i}{\partial a_i}\right\vert_{(\hat{a}^\epsilon,\hat{b%
}^\epsilon)}-\hat{\lambda}_i+\hat{\mu}_i=0, \\
&\hat{\lambda}_i(\hat{a}_i^\epsilon-x_i^0)=0,  \notag \\
&\hat{\mu}_i\hat{a}_i^\epsilon=0.  \notag
\end{align}
Equation (\ref{focq1}) can be written as 
\begin{equation*}
\begin{split}
\left.-\frac{\partial u_i}{\partial x_i}\right\vert_{(\hat{a}^\epsilon,\hat{b%
}^\epsilon)}+\left.\frac{\partial u_i}{\partial y_i}\right\vert_{(\hat{a}%
^\epsilon,\hat{b}^\epsilon)}\frac{\hat{B}^\epsilon+\epsilon}{\hat{A}%
^\epsilon+\epsilon}\biggl(1-\frac{\hat{a}_i^\epsilon}{\hat{A}%
^\epsilon+\epsilon}\biggr) \\
+\sum_{j\in I^2}\gamma_i^j \left.\frac{\partial u_j}{\partial y_j}%
\right\vert_{(\hat{a}^\epsilon,\hat{b}^\epsilon)}\frac{\hat{b}_j^\epsilon}{%
\hat{B}^\epsilon+\epsilon}-\hat{\lambda}_i+\hat{\mu}_i=0.
\end{split}%
\end{equation*}
Note that the summation over $j\in I^2$ is non-negative as the internal
utility functions $u_j$ are increasing, by Assumption 2, and $\gamma_i^j\geq
0$, by the assumption of the theorem, for each $j\in I^2$. Furthermore, the
multiplier $\hat{\mu}_i$ is non-negative, by the Kuhn-Tucker Theorem, and $%
\frac{\hat{B}+\epsilon}{\hat{A}+\epsilon}\geq C$, by Lemma 2. But then, from
the previous equation we can derive the following inequality which must hold in equilibrium 
\begin{equation}  \label{inequalityA}
\left.-\frac{\partial u_i}{\partial x_i}\right\vert_{(\hat{a}^\epsilon,\hat{b%
}^\epsilon)}+\left.\frac{\partial u_i}{\partial y_i}\right\vert_{(\hat{a}%
^\epsilon,\hat{b}^\epsilon)}C\biggl(\frac{\hat{A}^\epsilon-\hat{a}%
_i^\epsilon+\epsilon}{\hat{A}^\epsilon+\epsilon}\biggr)-\hat{\lambda}_i\leq
0.
\end{equation}
Suppose now that $a_i\rightarrow 0$. Then, we have that $\frac{\partial u_i}{%
\partial y_i}\rightarrow\infty$, as $\frac{\partial u_{i}(x_{i},0)}{\partial
y_{i}}=\infty$ by Assumption 3, $\frac{\hat{A}^\epsilon-\hat{a}%
_i^\epsilon+\epsilon}{\hat{A}^\epsilon+\epsilon}\rightarrow 1$, and $\hat{%
\lambda}_i^1=0$, as constraint $(i)$ is not binding for sufficiently small $%
a_i$. Furthermore, $\frac{\partial u_i}{\partial x_i}$ has an upper bound,
as $x_i>0$ and $u_i$ is continuously differentiable in the interior of the
consumption set by Assumption 2. But then, there exists an $\alpha>0$,
independent of $\epsilon$, such that the left hand side of equation (\ref%
{inequalityA}) is positive for each $a_i\in[0,\alpha]$. Hence, since the
inequality (\ref{inequalityA}) must hold in equilibrium, $\hat{a}%
_i^\epsilon>\alpha$, and, \textit{a fortiori}, $0<\alpha<\hat{A}^\epsilon$,
for each $\epsilon\in(0,1]$. We now consider the case in which there exists
an agent $i\in I^2$ who satisfies Assumption 3. Then, the previous argument
leads, \textit{mutatis mutandis}, to $0<\beta<\hat{B}^\epsilon$, for each $%
\epsilon\in(0,1]$.
\end{proof}

\bigskip

We can now prove the existence theorem.

\bigskip

\begin{proof}
Consider a sequence of $\{\epsilon_l\}_l$ converging to $0$. By Lemma 1, in
each perturbed game there exists an equilibrium. Then, we can consider a sequence of equilibria $\{(\hat{a}^{\epsilon_l},\hat{b}^{\epsilon_l})\}_l$. Since $%
S $ is compact and $\frac{\hat{B}^{\epsilon_n}+\epsilon_n}{\hat{A}%
^{\epsilon_n}+\epsilon_n}\in [C,D]$, by Lemma 2, we can pick a subsequence
of $\{(\hat{a}^{\epsilon_l},\hat{b}^{\epsilon_l})\}_l$ that converge to $(%
\hat{a},\hat{b})$ such that $(\hat{a},\hat{b})\in S$ and $\frac{\hat{B}}{%
\hat{A}}\in[C,D]$. But then, the strategy profile $(\hat{a},\hat{b})$ is a
point of continuity of payoff functions and then it is an equilibrium of $%
\Gamma(\gamma)$. The result of Lemma 3 implies that $%
\hat{A}>0$ or $\hat{B}>0$. But then, since $\frac{\hat{B}}{\hat{A}}\in[C,D]$%
, we can conclude that $\hat{A}>0$ and $\hat{B}>0$. Hence, $(\hat{a},\hat{b}%
) $ is a trade equilibrium.
\end{proof}

\end{document}